# CONSISTENT ESTIMATION IN LOGIT MODELS USING HISTORICAL CHOICES AS PRACTICAL CONSIDERATION SET

C. Angelo Guevara

Department of Civil Engineering, FCFM, University of Chile

Instituto Sistemas Complejos de Ingeniería (ISCI)

crguevar@ing.uchile.cl

## ABSTRACT

A key challenge in choice modeling lies in specifying the consideration set, the subset of alternatives that individuals actually evaluate when making choices, which is unobserved (latent) to the researcher. The classical *homo economicus* assumption posits that individuals assess the full universal set of alternatives, a behaviorally implausible premise. Practical options include directly asking individuals, which introduces behavioral biases; treating the consideration set as a latent construct, requiring full enumeration and strong identification assumptions; or relying on ad hoc heuristics that attempt to replicate how individuals form these sets or on non-parametric methods. Recently, some researchers have used historical choices as practical consideration set, an approach made increasingly feasible by the availability of passive data sources such as smartcards, mobile phone records, and scanner data. This article provides a formal demonstration of a sufficient condition, along with Monte Carlo evidence, showing that, under a Logit data-generating process with homogeneous choice probabilities across instances, defining a practical consideration set based on historical choices yields consistent parameter estimates. The demonstration is based on a reinterpretation of the sampling-of-alternatives theorem, viewing historical choices as draws from the true consideration set, and showing that under the stated assumptions, the uniform conditioning property holds. The article concludes by discussing the practical implications of this result and potential extensions to other modeling frameworks and assumptions.

## 1. INTRODUCTION

Choice modeling plays a key role in various fields, including transportation, marketing, health, and environmental studies. The conventional *homo economicus* paradigm in choice modeling assumes that decision-makers can evaluate all alternatives in the universal choice set, which is unrealistic from a cognitive demand standpoint, contradicting abundant behavioral science evidence. A more plausible approach to assume is that individuals assess only a small subset of alternatives in detail, forming what is known as the *consideration set*. However, this introduces a key challenge for researchers: since the consideration set is latent, it must be inferred for modeling purposes.

While theoretically rigorous approaches for constructing the consideration set exist, they are impractical for applied research. Conversely, practical methods currently in use lack robust theoretical foundations. This mismatch is problematic because incorrect assumptions about the consideration set can lead to inconsistent parameter estimates, ultimately undermining the validity of the entire modeling exercise (Villalobos and Guevara, 2021).

A first approach for inferring the consideration set is the direct elicitation of considered alternatives through surveys or interviews. While practical, but costly, this method suffers from recall bias, as respondents may inaccurately report the alternatives they evaluated (Nedungadi, 1990). Moreover, stated alternatives likely serve only as indicators of preferences. As a result, a two-step process model, where stated consideration precedes choice, would be misspecified (Horowitz and Louviere, 1995).

A second approach formally treats the consideration set as a latent variable within the choice process (Manski, 1977). While theoretically sound, this method becomes computationally intractable in settings with large choice sets (e.g., route choice), as it requires enumerating all $2^{|\Omega|}$ possible subsets, where $|\Omega|$ is the number of alternatives in the universal set. Identification in this case further demands either: (i) restrictive distributional assumptions, as in Swait & Ben-Akiva (1987); (ii) rather arbitrary behavioral constraints, as in Manzini & Mariotti (2014); or (iii) the availability of enriched datasets capturing choices under different frames, as in Gibbard (2021).

A third approach builds consideration sets by trying to approximate decision-makers' heuristic screening processes. In route choice modeling, practical methods often employ variants of k-shortest path algorithms (Prato, 2009). Similarly, Castro et al. (2013) propose selecting alternatives endogenously through attribute-based thresholds. However, trying to mimic decision-makers' heuristics, although sensate, is ad-hoc and unverifiable and, like direct elicitation, may lead to model misspecification.

A fourth approach adopts nonparametric identification strategies. For instance, Barseghyan et al. (2021) assume only a minimum choice set size and characterizes the sharp identification region of preferences using conditional moment inequalities. Although this framework is flexible and accommodates arbitrary dependence between preferences and choice sets, it generally delivers partial identification rather than point estimates. A final approach to the problem constructs consideration sets directly from revealed preference data, using historical choice patterns of either the modeled individual or similar decision-makers (see, e.g., Jánošíkova et al., 2014; Kim et al., 2020; Raveau et al., 2011, 2014; Yap et al., 2020; Arriagada et al., 2026). The proliferation of passive data sources (such as mobile GPS traces, transit smart cards, payment transactions, retail scanner data, and loyalty program records) has significantly enhanced the empirical viability of this method.

This article provides a formal demonstration of a sufficient condition under which, using an individual’s historical choices to define a practical consideration set, yields consistent parameter estimates when the data-generating process is Logit and choice probabilities are homogeneous across instances. The proposed approach can work for any type of choice set formation process behind the consideration set. The demonstration is based on reframing historical choices as draws from the true consideration set via McFadden’s (1978) sampling of alternatives theorem and establishing that the uniform conditioning property holds under specific assumptions. This allows estimating a curtailed logit model, with historical choices as practical consideration set, without any further correction. Monte Carlo evidence confirms, illustrates and characterizes the consistency result under these conditions and shows failure when attributes vary across choice instances. The article concludes by discussing the practical implications of this result and potential extensions to other modeling frameworks and assumptions.

## 2. ESTABLISHING CONSISTENCY UNDER HISTORICAL CONSIDERATION SETS

### 2.1 Problem Set-up and Notation

Consider a random utility maximization (RUM) discrete choice data generating process (DGP) in which an individual $n$, within a sample of size $N$, chooses the alternative $i$ with the largest utility, within the consideration set $C_n$. The utility $U_{in}$ that $n$ obtains from $i$ can be written as the sum of a systematic part $V_{in}$ and a random error term $\varepsilon_{in}$, as shown in Eq. (1)

$$U_{in} = V_{in} + \varepsilon_{in} = V\left(x_{in}, \beta^*\right) + \varepsilon_{in}, \tag{1}$$

where $V_{in}$ depends on $x_{in}$ attributes and population parameters $\beta^*$.

Then, if $\varepsilon_{in}$ is distributed *iid* Extreme Value I (0, $\mu$) the probability that $n$ chooses alternative $i$ will correspond to the Logit model shown in Eq. (2). The scale $\mu$ parameter in Eq. (2) is inversely proportional to the variance of $\varepsilon_{in}$. To address its lack of identifiability, we normalize $\mu = 1$ for all subsequent analyses.

$$P_n\left(i \mid C_n\right) = \frac{e^{\mu V_{in}}}{\sum_{j \in C_n} e^{\mu V_{jn}}} \tag{2}$$

Now suppose that while the true consideration set $C_n$ is latent to the researcher, she can observe the individual making $R+1$ choices from it. A *homogeneous probability* is assumed, i.e that the choice probabilities remain unchanged across the $R+1$ choice instances. This also implies that the consideration set $C_n$, the attributes $x_{in}$, and the choice behavior reflected in $\beta^*$ remain the same across the $R+1$ instances, and that only the realizations of the $\varepsilon_{in}$ change, resulting in different choices, following the probability model shown in Eq. (2).

The researcher is interested in modeling the choice that occurs in instance $R+1$ to obtain consistent estimators of the model parameters $\beta^*$ and, in that sense, the previous $R$ choices are termed historical choices. To do this, she builds a practical consideration set $D_n$, as shown in Eq. (3), including all the alternatives $j$ that were chosen by individual $n$ in the $R+1$ instances

$$D_n = \left\{ j \in C_n \mid m_j = \sum_{r=1}^{R+1} y_{jn}^{(r)} > 0 \right\}, \tag{3}$$

where $y_{jn}^{(r)}$ equals *1* if alternative $j$ is chosen by individual $n$ in instance $r$, and 0 otherwise; and $m_j$ denotes the number of times each alternative $j$ is chosen across the $R+1$ instances by individual $n$. The index $n$ is omitted in $m_j$ to save notation. Note that $\left|D_n\right|$, the cardinality of $D_n$, can range from 1, when the same alternative is chosen $R+1$ times, to $min(R+1, \left|C_n\right|)$, when all selected alternatives are unique or the consideration set is recovered. Note also that, by design, $D_n \subseteq C_n \subseteq \Omega_n$ and $\left|D_n\right| \le \left|C_n\right| \le \left|\Omega\right|$. The practical consideration set includes all the historical choices, i.e. the alternatives chosen in the past *R* instances, and adds the alternative chosen in instance *R+1*, if it was not previously included.

The following proposition formalizes the main result of the paper:

***Proposition 1: Consistency under Historical Consideration Sets***

*Consider a random utility maximization (RUM) model where a sample of N individuals n make R+1 choices from a latent consideration set* $C_n$. *The choices follow a Multinomial Logit (MNL) structure with true parameter vector* $\beta^*$. *Let the observed practical consideration set* $D_n$ *be defined as in Eq. (3).*

*Assume the following:*

*(A1) i.i.d. Errors and Exogeneity: The unobserved utility components* $\varepsilon_{inr}$ *are independently and identically distributed (i.i.d.) Extreme Value Type I across all alternatives j, individuals n, and choice instances r* $\in \{1, \ldots, R+1\}$. *Furthermore, the errors are assumed to be strictly independent of the observed attributes* $x_{inr}$.

*(A2) Homogeneity: Choice probabilities, and by extension* $C_n = C_{nr}$ , *attributes* $x_{inr} = x_{in}$ *and parameters* $\beta^*$, *remain constant across the R+1 choice instances. Then:*

*Estimating a curtailed Logit model for the choice* $i^{(R+1)}$ *observed in instance R+1, using* $D_n$ *as the practical consideration set, as shown in Eq.(4), yields a consistent estimator of* $\beta^*$ *as N→ ∞, and R→ ∞.*

$$\pi\left(i^{(R+1)} \mid D_n\right) = \frac{e^{V_{in}}}{\sum_{j \in D_n} e^{V_{jn}}} \tag{4}$$

*Moreover, the estimator is asymptotically normal, with variance–covariance matrix consistently estimated by the robust (sandwich) estimator.*

Monte Carlo evidence presented in Section 3 further suggests that this consistency result may hold even for small fixed values of *R* (e.g., $R \geq 5$) under the homogeneity assumption. However, a formal proof of exact fixed-*R* consistency remains an open problem, as it is discussed further in the Conclusion.

### 2.2 From Sampling of Alternatives to Consideration Sets.

To describe the proposed approach, we first need to revisit McFadden's (1978) sampling-of-alternatives problem for modeling residential location choice. It should be remarked that this problem is structurally different from the consideration set problem studied in the present article, although sometimes they are confounded (see e.g., Hicks et al, 2020). In the sampling-of-alternatives problem it is assumed that the true consideration set is the universal set $\left(C_n = \Omega_n\right)$, which can be fully assessed by rational decision-makers and is also known by the researcher. In this case, the sole limitation is that the researcher cannot feasibly handle the universal consideration set, owing to computational and data collection burdens. The solution proposed by McFadden (1978) to solve this problem involves constructing a reduced set $S_n \subseteq C_n$ via a sampling protocol conditioned on the chosen alternative, enabling feasible estimation while preserving consistency.

Formally, define $\pi(S_n \mid j)$ as the conditional probability that the researcher would sample a reduced set $S_n$, given that alternative $j$ was chosen by individual $n$. For example, if the true consideration set has 10 alternatives, a sampling protocol could be to draw the chosen alternative and then randomly draw without replacement, and with equal probability, 2 non-chosen alternatives to end up with a set $S_n$ of only 3 alternatives. Under this setting, using combinatorics it can be easily shown that $\pi(S_n \mid j) = \binom{9}{2}^{-1} = 1/36$ .

McFadden (1978) showed that, under a Logit DGP and a given sampling of alternatives protocol $\pi(S_n \mid j)$, consistent estimators of the population parameters $\beta^*$ can be obtained by maximizing a pseudo-likelihood using the probabilities of choice shown in Eq. (5).

$$\pi(i \mid S_n) = \frac{e^{V_{in} + \ln \pi(S_n|i)}}{\sum_{j \in S_n} e^{V_{jn} + \ln \pi(S_n|j)}} \tag{5}$$

The advantage of the pseudo-likelihood described in Eq. (5) is that it depends only on the alternatives of the reduced set $S_n \subseteq C_n$, transforming a problem of possibly millions of alternatives into one that has only a few. Besides, the resulting model has a closed form that corresponds to a simple Logit model with a *sampling correction* $\ln \pi(S_n \mid j)$ by alternative that only depends on the sampling protocol, which is defined by the researcher. This result holds thanks to the IIA property of the Logit model but was extended to more flexible models such as MEV (GEV), Logit Mixture and RRM models, by Guevara and Ben-Akiva (2013 a,b) and Guevara et al. (2016), respectively.

Besides, for some sampling protocols, the sampling correction $\ln \pi(S_n \mid j)$ is independent of the alternative chosen by the individual, i.e $\ln \pi(S_n \mid j) = \ln \pi(S_n \mid j') = \lambda_n \quad \forall j, j' \in S_n$ a condition known as the **uniform conditioning property**. In such cases, the sampling correction cancels out in Eq. (5), yielding a pseudo-likelihood equivalent to that of curtailed simple Logit model, with the sampled choice set $S_n \subseteq C_n$.

$$\pi(i \mid S_n) = \frac{e^{V_{in} + \ln \pi(S_n|i)}}{\sum_{j \in S_n} e^{V_{jn} + \ln \pi(S_n|j)}} \frac{e^{V_{in} + \lambda_n}}{\sum_{j \in S_n} e^{V_{jn} + \lambda_n}} = \frac{e^{V_{in}} e^{\lambda_n}}{\sum_{j \in S_n} e^{V_{jn}} e^{\lambda_n}} = \frac{e^{\lambda_n} e^{V_{in}}}{e^{\lambda_n} \sum_{j \in S_n} e^{V_{jn}}} = \frac{e^{V_{in}}}{\sum_{j \in S_n} e^{V_{jn}}} \tag{6}$$

This uniform conditioning property holds, for example, in the previously described case where 2 non-chosen alternatives out of the remaining 9 were randomly drawn without replacement, resulting in $\ln \pi(S_n \mid j) = 1/36$ for all *j*. For other sampling protocols, however, the uniform conditioning property may not hold, and the sampling correction may therefore not be canceled out, potentially leading to the need of substantial adjustments. Examples of these latter cases are discussed by McFadden (1978), Ben-Akiva and Lerman (1985, Ch. 9), Ben-Akiva (1989), among others. In this context, statements that estimating a multinomial logit model using a true subset of the unobserved choice set yields consistent estimators can be misinterpreted if taken without qualification. In general, a sampling correction is required and cancels out only when the uniform conditioning property holds.

While the sampling alternatives problem is structurally different from the consideration set problem, in what follows we will use the former to develop a novel practical solution for the latter. The key is that, while the source of the subset in both problems differs (researcher-led sampling vs. agent-led historical choice), the mathematical properties of the resulting likelihood are mathematically equivalent under the homogeneity assumption.

Reconsider the historical consideration set problem depicted in Subsection 2.1. As stated before, it consists in modeling the choice in instance *R+1* using as consideration set the alternatives that were chosen in the *R+1* instances. Thus, while the consideration set $C_n$ is latent to the researcher, for the sake of the choice modeled at instance *R+1*, all historical (past) *R* choices can be seen as draws from the latent $C_n$, with probability $P_n(i \mid C_n)$, under the homogeneity assumption.

Thus, the DGP for the historical consideration set problem defined in Section 2.1, to model the choice at instance R+1, can be interpreted as a sampling-of-alternatives problem, where: (i) the universal set corresponds to the consideration set $C_n$; (ii) the *R* historical choices are understood as draws from $C_n$, with probability $P_n(i \mid C_n)$, as shown in Eq. (2), and (iii) where the reduced set is defined as $S_n \equiv D_n = \left\{ j \in C_n \mid m_j = \sum_{r=1}^{R+1} y_{jn}^{(r)} > 0 \right\}$.

## 2.3 Proof of Proposition 1

By framing the consideration set problem (Section 2.1) as a sampling of alternatives problem, we can directly apply McFadden's (1978) theorem (Eq. 5). It follows that consistent estimators of the population parameters $\beta^*$ are obtainable by maximizing a pseudo-likelihood based on the choice probabilities defined in Eq. (7).

$$\pi\left(i^{(R+1)} \mid D_n\right) = \frac{e^{V_{in} + \ln \pi\left(D_n \mid i^{(R+1)}\right)}}{\sum_{j \in D_n} e^{V_{jn} + \ln \pi\left(D_n \mid j^{(R+1)}\right)}} \tag{7}$$

Given Eq. (7), the proof of Proposition 1 reduces then to determining the appropriate sampling correction $\ln \pi(D_n \mid j)$ and verifying that the uniform conditioning property holds. Under this setting, the sampling protocol in this case is equivalent to importance sampling with replacement, in which the sampling probability of each alternative corresponds to the choice probability $P_n\left(i \mid C_n\right)$ in each instance $r$. Note that, although the sampling process is with replacement, which reflects on counts of choices by alternative $m_j$, the observed object collapses to a set $D_n$, as shown in Eq. (3). The solution builds on a closely related problem developed by Ben-Akiva (1989), for the case of importance sampling with replacement as an addition to Ben-Akiva and Lerman (1985, Ch. 9).

Consider the conditional probability $\pi\left(D_n \mid i^{(R+1)}\right)$ of generating the set $D_n$, given alternative $i$ is chosen in instance $R$+1, as shown in Eq. (3). Under the homogeneous probability assumption, and letting $m_j$ denote the number of times alternative $j$ is chosen across the $R+1$ instances, $\pi\left(D_n \mid i^{(R+1)}\right)$ is given by the multinomial probability mass function shown in Eq. (8), over the count vector $M_{(i)}=\left(m_1,\cdots,m_i-1,\cdots,m_j,\cdots,m_{|D_n|}\right)$.

$$\pi\left(D_n \mid i^{(R+1)}\right)=\frac{R!}{\left(m_i-1\right)!\prod_{\substack{j\in D_n \\ j\neq i}} m_j!} P_n\left(i \mid C_n\right)^{m_i-1} \prod_{\substack{j\in D_n \\ j\neq i}} P_n\left(j \mid C_n\right)^{m_j} \tag{8}$$

Note that since the derivation is conditioned on the choice at instance $R+1$ (alternative $i$), the count for that specific alternative is reduced by one in the vector $M_{(i)}$ representing instances $1$ through $R$. Furthermore, while the dimension of $M_{(i)}$ corresponds to the number of unique alternatives in the historical set $D_n$, its indices may not align with those of the consideration set $C_n$ . However, there exists a one-to-one correspondence (or perfect concordance) between the indices of $M_{(i)}$ and the specific alternatives identified in $C_n$, ensuring the multinomial probability in Eq. (8) is correctly attributed.

The expression in Eq. (8) can be substantially simplified when deriving the sampling correction required in Eq. (7). To this end, we first complete both product operators to include alternative $i$ and, to preserve the value of Eq. (8), multiply by the corresponding inverse term $m_i / P_n\left(i \mid C_n\right)$, yielding the expression on the center of Eq. (9). We then group all terms that do not vary across alternatives in $K_{D_n}$, for each individual $n$, as shown on the right-hand side of Eq. (9). This term cancels out in the choice probability in Eq. (7) and can therefore be later ignored.

$$\pi\left(D_n \mid i^{(R+1)}\right)=\frac{m_i}{P_n\left(i \mid C_n\right)} \underbrace{\left(\frac{R!}{\prod_{j\in D_n} m_j!} \prod_{j\in D_n} P_n\left(j \mid C_n\right)^{m_j}\right)}_{K_{D_n}}=\frac{m_i}{P_n\left(i \mid C_n\right)} K_{D_n}. \tag{9}$$

However simple in form, after ignoring $K_{D_n}$, Eq. (9) is still not practical, as it depends on the choice probabilities $P_n\left(i \mid C_n\right)$, which in turn depend on the unknown parameters $\beta^*$ and the latent consideration set $C_n$.

Nevertheless, as shown in the appendix (Proposition 2), $\hat{P}_n\left(i \mid C_n\right)$ in Eq. (10) is a consistent (in $R$) estimator of the choice probability $P_n\left(i \mid C_n\right)$.

$$\hat{P}_n\left(i \mid C_n\right)=\frac{m_i}{\sum_{j \in D_n} m_j}=\frac{m_i}{R+1} \quad , \tag{10}$$

Thus, Since $\hat{P}_n\left(i \mid C_n\right) \xrightarrow{p} P_n\left(i \mid C_n\right)$ and the mapping is continuous, the Slutsky theorem (Ben-Akiva and Lerman, 1985), implies that $\tilde{\pi}\left(D_n \mid i^{(R+1)}\right)$ in Eq. (11) will be a consistent estimator of $\pi\left(D_n \mid i^{(R+1)}\right)$ as $R \to \infty$.

$$\tilde{\pi}\left(D_n \mid i^{(R+1)}\right)=K_{D_n} \frac{m_i}{\frac{m_i}{R+1}}=\left(R+1\right) K_{D_n}, \tag{11}$$

Then by the Slutsky and McFadden's (1978) theorems, replacing $\pi\left(D_n \mid i^{(R+1)}\right)$ by $\tilde{\pi}\left(D_n \mid i^{(R+1)}\right)$ in Eq. (7) will yield consistent estimators of the model parameters $\beta^*$ as $R \to \infty$ and $N \to \infty$. Furthermore, since $\tilde{\pi}\left(D_n \mid i^{(R+1)}\right)$ is invariant in $i$, it cancels outs, implying that a curtailed Logit model in Eq. (4) yields consistent estimates for the historical consideration set problem under this two-limit argument, proving the first part of Proposition 1. Whether exact consistency holds at any fixed $R \geq 1$, as suggested numerically later in Section 3 for the homogeneous-attribute case, remains an open question.

Note that $\tilde{\pi}\left(D_n \mid i^{(R+1)}\right)$ from Eq. (11), is invariant across $i$ for any fixed $R \geq 1$ because the estimator $\hat{P}_n\left(i \mid C_n\right)$ in Eq. (10) cancels $m_i$ when replaced in Eq. (9), regardless of the value of $R$. This exact cancellation, however, is not sufficient for consistency of $\beta^*$, because $\tilde{\pi}\left(D_n \mid i^{(R+1)}\right)$ is a substitute for the true sampling correction $\pi\left(D_n \mid i^{(R+1)}\right)$ given in Eq. (9). The correction and its approximation coincide only as $\hat{P}_n\left(i \mid C_n\right) \xrightarrow{p} P_n\left(i \mid C_n\right)$ that is, as $R \to \infty$ (Proposition 2). It is this convergence, and not the cancellation step itself, that requires $R \to \infty$. In that sense, it can be said that the uniform conditioning property holds only "asymptotically", while the approximated correction $\tilde{\pi}\left(D_n \mid i^{(R+1)}\right)$ cancels out for any value of $R$.

Finally, given that $\hat{P}_n\left(i \mid C_n\right)$ is also unbiased (see appendix), the two-stage approach of Train (2009, Section 10.5) can be used to establish that the estimators of the curtailed model in Eq. (4), i.e. when the sampling correction is ignored, are asymptotically normal, with variance–covariance characterized by the "robust" (sandwich) estimator, as reported in Eq. (12-13).

$$\hat{\beta} \overset{a}{\sim} \text{Normal}\left(\beta^*, \Omega/N\right), \tag{12}$$

where $\Omega$ can be calculated with the "sandwich estimator" as

$$\hat{\Omega} = \left[\frac{\partial^2 \ln \hat{\pi}(\hat{\beta} \mid D)}{\partial \beta \, \partial \beta'}\right]^{-1} \left[\sum_{n=1}^{N} \frac{\partial \ln \hat{\pi}_n(\hat{\beta} \mid D)}{\partial \beta} \frac{\partial \ln \hat{\pi}_n(\hat{\beta} \mid D)}{\partial \beta'}\right] \left[\frac{\partial^2 \ln \hat{\pi}(\hat{\beta} \mid D)}{\partial \beta \, \partial \beta'}\right]^{-1} \quad (13)$$

## 3. MONTE CARLO EXPERIMENT

This section presents a Monte Carlo experiment designed to illustrate the consistency result derived in Section 2 for the historical consideration set problem, thereby confirming its validity, and illustrating the impact of violating the assumption of homogeneous probability.

The experiment simulates a Logit choice model with 1,000 individuals, each facing a universal choice set of 100 alternatives described by three variables: *a*, *b* and *p*, where the third can be seen as the "price" and the other as positively valued attributes of each alternative. Attribute *b* is generated as standard normal, whereas *a* is generated as the sum of a standard normal and a term $j/|\Omega|$ that grows linearly with the index *j* of the alternative. Attribute *p* is generated as *b* minus *a* plus a standard normal. Thus, *a* depends on the alternative's index, and *p* is correlated with both *a* and *b*.

Choices are modelled using a Logit model with linear utility specification and coefficients $\beta_a^*$ = 0.5, $\beta_b^*$=1 and $\beta_p^*$=-2, for *a*, *b* and *p* respectively. The consideration set for each individual *n* includes 10 alternatives per individual, which are determined via a Logit model with linear utility and coefficients 1, 1 and -1 for *a*, *b* and *p*, respectively. A variant of the experiment was also conducted by generating the consideration set using an elimination-by-aspects approach (Tversky 1972), yielding qualitatively the same results, confirming that the proposed approach can work for any type of choice set formation process behind the consideration set.

The experiment is divided into two cases: one with fixed attributes (fulfilling the homogeneity assumption), and another with variable attributes (violating it). For each case, six models are estimated. The first model uses all 100 alternatives (All) as the practical consideration set, to illustrate the impact of a misspecification of the consideration set. The second model uses the true consideration set (True), which serves as a benchmark for the best possible performance. The other models use practical consideration sets built from different numbers of instances, *R*=5,10, 20, and 40, to study the impact of *R*. The experiment is replicated 100 times, from which various statistics are computed and reported in Figure 1 and Tables 1 and 2, which are complemented with Figure 2 and Table 3, reported in the appendix.

Figure 1 presents boxplots depicting the distribution of the ratio of the estimated coefficients $\hat{\beta}_a / \hat{\beta}_b$ across the 100 replications of the experiment for the six models. On the left are presented the results for the case of fixed attributes and historical sets across all *R+1* instances, fulfilling the homogeneity assumption. On the right are presented the results of the case of variable attributes and historical sets, which are regenerated for each choice instance, breaking the homogeneity assumption. The population value of the ratio $\beta_a^* / \beta_b^* = 0.5$ is highlighted by a dashed line. The average of each empirical distribution is highlighted with a bold line over each boxplot. These boxplots highlight the empirical bias and efficiency achieved by different models.

Figure 1 is complemented with the results shown in Tables 1 and 2. Table 1 depicts the percentual mean bias, the standard error, and the p-value of a t-test for the hypothesis that the ratio $\beta_a^* / \beta_b^*$ is equal to the population value of 0.5, for each model (by row) and case (fixed attributes on the left and variable attributes on the right). Additionally, Table 2 reports the coverage rate, defined as the percentage of the sample with a historical consideration set of a given size $|D_n|$ (from 1 to 10), as indicated by column. The results are presented by row for each case (fixed and variable attributes) and model (only $R$=5,10, 20, and 40 in this case). Thus, for example, in the fixed attributes experiment with $R$=5, 34% of individuals have a historical choice set of size 3. In contrast, in the variable-attributes experiment with $R$=40, 75% of individuals have a historical choice set of size 10. In both cases, these correspond to the modal values for each experiment and can be compared to the true consideration set cardinality of 10.

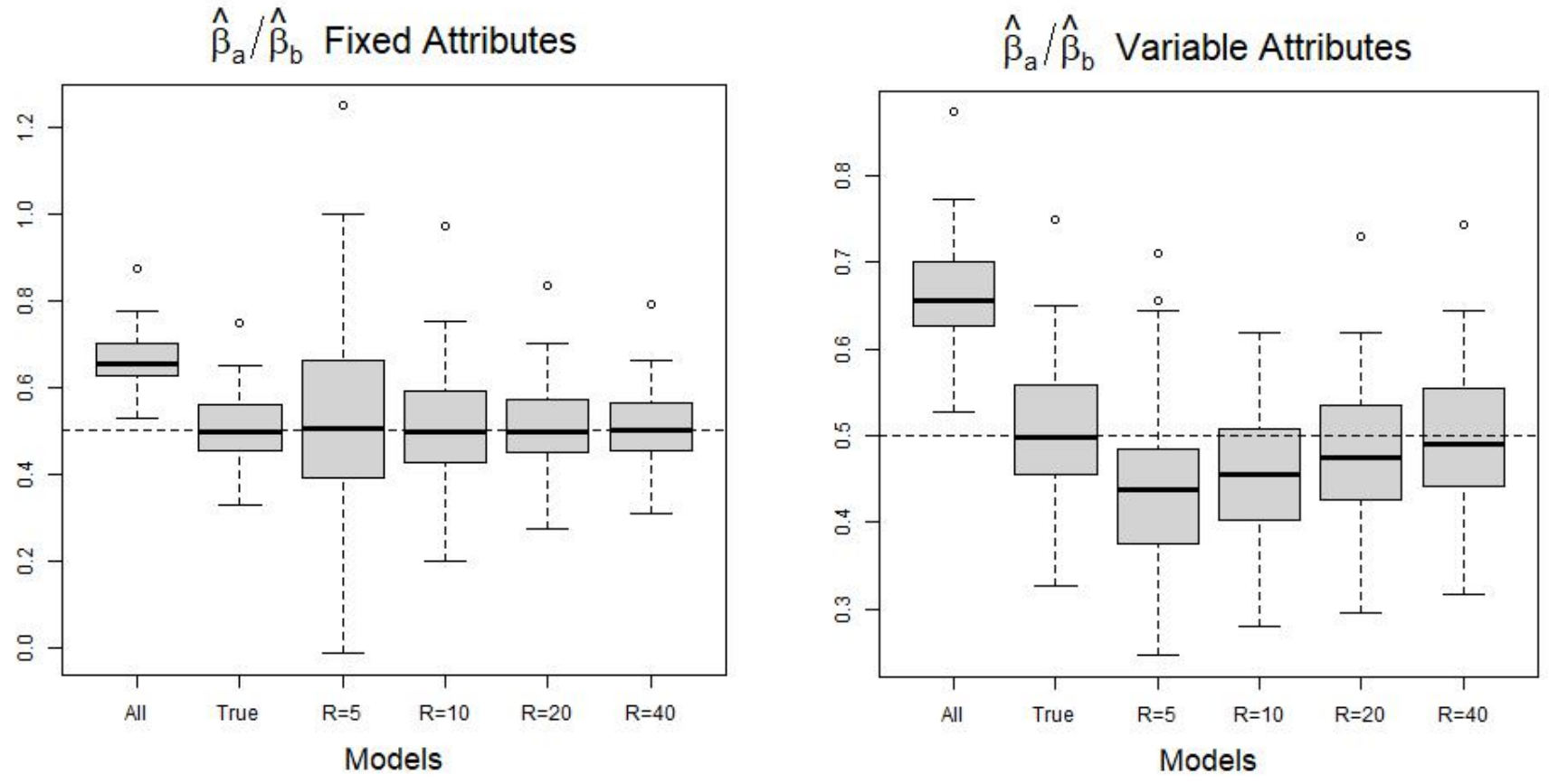

Figure 1. Estimators Under Different Assumptions and Models

The *All* model, both for Figure 1 left and right, shows that when the wrong consideration set is used (in this case the universal set) a significant bias may arise. In both cases all the mass of the empirical distribution of $\hat{\beta}_a / \hat{\beta}_b$ is above the true value of 0.5, with an average bias of 32% and a p-value <0.01%, as shown in Table 1. The *True* model, both for Figure 1, left and right, shows that the experiment is well specified, in the sense that the empirical distribution is centered on the population value, with a bias of 0.77% and a p-value 60%, as shown in Table 1. This model also works as a benchmark for the best possible efficiency (shorter standard error of 0.0073, and narrower boxplot) that could be achieved.

**Table1. Bias and p-value of Recovering Population Values by R**

| | Fixed Attributes | | | Variable Attributes | | |
|---|---|---|---|---|---|---|
| Models | % bias | s.e | % p-value | % bias | s.e | % p-value |
| All | 32 | 0.0056 | <0.01 | 32 | 0.0056 | <0.01 |
| True | 0.77 | 0.0073 | 60 | 0.77 | 0.0073 | 60 |
| R=5 | 4.6 | 0.020 | 27 | -13 | 0.0093 | <0.01 |
| R=10 | 1.8 | 0.013 | 49 | -8.8 | 0.0074 | <0.01 |
| R=20 | 0.78 | 0.0096 | 69 | -4.8 | 0.0075 | 0.20 |
| R=40 | 0.77 | 0.0083 | 65 | -1.0 | 0.0074 | 50 |

In turn, the models with $R$ = 5, 10, 20, and 40 in Figure 1 present a contrasting picture when comparing the case when attributes are fixed (left) and when they vary (right), breaking the homogeneous probability assumption. On the one hand, when the attributes are fixed across choice instances (i.e. choice probabilities are homogeneous), as shown in the left-hand panel of Figure 1, all empirical distributions are centered around the population value (0.5), even with as few as 5 instances, confirming the validity of the demonstration presented in Section 2. Consequently, Table 1 shows p-values of 27% and larger for all values of $R$ analyzed. The only aspect that changes as $R$ increases is the efficiency of the estimators, reflected in narrower boxplots in Figure 1, larger p-values, and smaller standard errors in Table 1, which increasingly approach the performance achieved by the benchmark, i.e. using the true consideration set.

On the other hand, when the attributes vary across choice instances (i.e. choice probabilities are not homogeneous), as shown in the right-hand panel of Figure 1, only when $R$=40 the empirical distribution is centered around the population value of 0.5. For all other values of $R$ the bias is large and, as shown in Table 1, the p-values are significantly smaller than the typical threshold of 5%. This illustrates the potentially large impact of violating the homogeneous probability assumption.

Besides, Table 2 shows that the empirical bias in the variable attributes case eventually disappears only due to exhaustion, that is, the near-complete recovery of the true consideration set as the number of realizations increases. This is supported by the fact that the true consideration set has a cardinality of 10, and that the test for recovery of the population parameters at the 5% significance level is passed only for R=40, when 75% of the sample has a historical set equal to the true consideration set and 97% (75% + 22%) is missing at most one alternative.

**Table 2. Consideration Set Coverage(*) Rate % by R**

| | $R/\lvert D_n\rvert$ | 1 | 2 | 3 | 4 | 5 | 6 | 7 | 8 | 9 | 10 | Total |
|---|---|---|---|---|---|---|---|---|---|---|---|---|
| Fixed Attributes | 5 | 16 | 30 | 34 | 17 | 4 | 0 | 0 | 0 | 0 | 0 | 100 |
| | 10 | 9 | 18 | 27 | 26 | 14 | 5 | 1 | 0 | 0 | 0 | 100 |
| | 20 | 5 | 10 | 16 | 26 | 25 | 12 | 5 | 1 | 0 | 0 | 100 |
| | 40 | 3 | 5 | 10 | 17 | 22 | 21 | 15 | 6 | 1 | 0 | 100 |
| Variable Attributes | 5 | 0 | 1 | 8 | 33 | 46 | 12 | 0 | 0 | 0 | 0 | 100 |
| | 10 | 0 | 0 | 0 | 2 | 8 | 30 | 39 | 17 | 4 | 0 | 100 |
| | 20 | 0 | 0 | 0 | 0 | 0 | 2 | 8 | 29 | 43 | 18 | 100 |
| | 40 | 0 | 0 | 0 | 0 | 0 | 0 | 0 | 3 | 22 | 75 | 100 |

(*) Percentage of the sample with a historical consideration set of size $\lvert D_n\rvert$ for each experiment.

As a complement to the analysis in Figure 1 and Table 1, Table 3 (Appendix) provides a summary of the estimation results for all parameters across the 100 simulations and six models. It is important to note that the recovery of the underlying model should not be assessed based on individual coefficients in isolation. In discrete choice modeling, the absolute scale is not identified; instead, consistency is typically evaluated through the ratios of coefficients, as reported in Figure 1 and Table 1. Nevertheless, the full set of results can also provide useful insights into what is occurring in practice.

Interestingly, the results in Table 3 seem to suggest that the fixed-attribute case performs worse than the variable-attribute case. For R=5, the true parameter $\beta_a^*$ = 0.5 is estimated at 0.185 in the fixed case, compared to 0.436 in the variable case, with the latter much closer to its true value. This discrepancy arises because the practical consideration set for fixed attributes is significantly smaller than for variable attributes (see Table 2). This reduction in information increases the relative error variance of the choice model, leading to a smaller model scale and, consequently, lower nominal parameter values. The fact that the fixed attributes case does perform better than the variable case can be confirmed in Table 3 by looking at the ratios of the mean estimators $\hat{\beta}_a / \hat{\beta}_b$, which is almost perfectly recovered for the first case, but not for the second. The other interesting value that is reported in Table 3 is the fact that the RMSE is significantly reduced and the number of historical choices grows from 5 to 40, confirming its impact on efficiency.

To further explore the properties of the estimator, Figure 2 (Appendix) presents an analysis of the "empirical consistency" of the estimator of the parameter ratio $\hat{\beta}_a / \hat{\beta}_b$ for R=5. By examining a single realization across increasing sample sizes ($N$ ranging from 1,000 to 121,000), the 95% confidence intervals (estimated with the delta method) reveal a clear divergence between the two settings. In the fixed-attribute case, the true ratio is recovered with precision. It can be seen that as the sample size increases, the estimates converge toward the true value with progressively smaller variances, confirming empirical consistency. Conversely, this convergence toward the true ratio is not observed in the variable-attribute case, which shows a negative bias. These results suggest that while the fixed-attribute setting may have a smaller practical consideration set, affecting the scale, the trade-offs between attributes remain identifiable, whereas the non-homogeneity inherent in the variable case introduces a more fundamental challenge to parameter recovery.

## 4. CONCLUSION

This article formally proves that if choice probabilities are homogeneous across choice instances of a Logit model, using historical choices to define consideration sets ensures consistent parameter estimation. Four key implications of this result are discussed in this final section: (i) the plausibility of the homogeneity assumption, (ii) the feasibility of obtaining the required historical choice data, (iii) the robustness of the result to real-world assumption violations, and (iv) potential extensions to heterogeneous settings and non-logit models.

The homogeneous probability assumption requires that consideration set $C_n$, the attributes $x_{in}$, and the preference parameters $\beta^*$ remain constant across all $R$+1 choice instances. While this may hold for classical models, where $C_n$ and $\beta^*$ are typically assumed stable, a more realistic approach could incorporate learning, habit formation (e.g., Henríquez-Jara and Guevara, 2025; Henríquez-Jara et al., 2025), or attribute variation. For example, transportation settings often involve fluctuating travel times, violating fixed $x_{in}$, whereas scanner data may sustain this assumption temporarily if prices remain unchanged.

Regarding the feasibility of the method, while collecting historical data remains a significant challenge for classical survey-based sources, it has become increasingly viable with the rise of passive data streams. Technologies such as smartcard transactions, telecommunication records, and scanner data provide the high-frequency longitudinal observations necessary to reconstruct historical choice sets with high precision. While many challenges still need to be solved to transform passive data sources into useful information for this regard, there is a growing literature in the topic that already has great advances. For example, Arriagada et al (2022, 2025, 2026) can effectively recover historical public transportation route choices by combining Smart Card, GPS and GTFS data from buses.

One line of future research involves extending these consistency results to settings where the homogeneity assumption is relaxed, as this restriction is often untenable in empirical applications and, as illustrated in Figure 1, can induce significant estimation biases. The fundamental challenge lies in the fact that when choice probabilities vary across instances, the sampling correction $\ln \pi\left(D_n \mid i^{(R+1)}\right)$ can no longer be expressed in the closed-form multinomial PMF of Eq. (8). Instead, the counts follow a non-homogeneous multinomial distribution (Johnson et al., 1997), where the specific functional form of the correction remains analytically elusive. Our Monte Carlo experiments confirm that, in such non-homogeneous cases, the Uniform Conditioning Property is generally violated, necessitating more complex identification strategies. However, it may be possible to mitigate this by identifying specific data transformations where the correction factor remains constant across choice instances. Under such conditions, provided that such transformations do exist, a fixed-effects Logit framework (Stammann et al., 2016) could potentially yield consistent estimates by jointly modeling all $R$+1 instances. While the empirical survey by Crawford et al. (2021), building on Chamberlain (1980), provides a practical basis for this perspective, formal theoretical proof remains an open challenge.

Regarding the impact of relaxing the homogeneity assumption, which may be difficult to sustain in transportation settings, empirical evidence from Arriagada et al. (2026) suggests that the practical impact may be limited or, at the very least, that the proposed method performs favorably relative to other feasible alternatives. Arriagada et al (2026) used a cohort-based approach, using choices from similar travelers (same OD pair and period) rather than individual's own historical data, yielding superior results to existing methods in public transport route choice modeling. Their three-week estimation/fourth-week validation design showed this method outperformed heuristic approaches, delivering better in-sample fit and more accurate out-of-sample predictions, even when stronger assumptions of inter-persons homogeneity were placed. These empirical successes suggest that leveraging historical choice data represents a preferred methodology for accurately capturing latent consideration sets in complex discrete choice environments."

Regarding a formal test for potential violations of the homogeneity assumption, and provided that a sufficiently large set of historical choices is available, one could in principle apply a chi-square test of homogeneity (see, e.g., Agresti, 2013). For instance one may partition the historical choices into two arbitrary groups, to end up with measures $m_{j1}$ and $m_{j2}$, with which to apply a chi-square test of homogeneity. However, in practice, it appears unlikely that the available data would be large enough to ensure adequate statistical power for such a test.

More generally, it should be noted that this manuscript provides only a sufficient condition for consistency, the homogeneity assumption. No formal proof is offered that violations of this assumption necessarily lead to inconsistency; rather, the evidence of that is based on the Monte Carlo case study. Interestingly, a slight variation of the case study, in which the mean of variable $a$ does not depend on the index of the alternative, exhibits what may be termed "empirical consistency," with virtually zero finite-sample bias as $N$ grows. Such a data-generating process can be interpreted as a case of fully unlabeled and indistinguishable alternatives. A formal proof of consistency for this and other special cases has been preliminarily explored by Guevara and Ben-Akiva (2026), although further analysis and a deeper examination of its practical implications are still ongoing.

Another line of research involves extending these results to non-Logit models where the IIA property is relaxed, a critical requirement for many empirical settings. In these contexts, the standard McFadden (1978) theorem (Eq. 5) no longer holds directly. However, as noted previously, McFadden's framework has been extended to MEV (GEV), Logit Mixture, and Random Regret Minimization (RRM) models by Guevara and Ben-Akiva (2013a,b) and Guevara et al. (2016), respectively. A primary challenge in applying these approaches here is that sampling alternatives in non-IIA contexts requires expansion factors to correct for truncated terms within the modeled probabilities. Current literature distinguishes between scenarios where the researcher can resample alternatives to construct these expansion factors and those where resampling is impossible. The former generally yields superior performance because, in the absence of resampling, the expansion factors remain dependent on the very choice probabilities that the researcher seeks to estimate, requiring they be approximated in practice. Since resampling is not feasible for the historical choice problem, where the observed sets are fixed by past behavior, this latter case holds. Consequently, a systematic exploration of approximation methods to mitigate this circular dependency represents a natural first step toward extending the consistency results of the present paper to non-Logit models. Such an extension would bridge the gap between the theoretical robustness of historical consideration sets and the flexible substitution patterns required in complex discrete choice applications.

A further open question concerns whether the consistency result of Proposition 1 holds exactly for any fixed $R \geq 1$, rather than only asymptotically as $R \to \infty$. The present proof relies on replacing the true choice probability $P_n\left(i \mid C_n\right)$ with the empirical frequency $\hat{P}_n\left(i \mid C_n\right)$, a substitution justified via Slutsky's theorem only as $R \to \infty$ (Proposition 2, Appendix). Our Monte Carlo results (Table 1, Figure 2) indicate that, under the homogeneity assumption, the empirical bias is already negligible at small fixed $R$ (e.g., $R = 5$), suggesting that a stronger, exact fixed-$R$ result may hold. The study of this potential formal proof is left for future research.

Finally, several practical considerations for the application of the proposed method arise. One is that there seems to be a tradeoff between using a large $R$ to improve efficiency, by more closely approximating the true consideration set, and ensuring stability in choice conditions (e.g., infrastructure and preferences). A comprehensive assessment of this tradeoff ultimately requires extensive empirical analysis. A related challenge concerns the use of this method, primarily intended for estimation, for forecasting applications or in contexts where no historical choices are available. In such cases, this limitation can be mitigated using the hybrid approach proposed by Arriagada et al. (2026), which combines parameters estimated from historical data with heuristic choice set generation to accommodate new or evolving contexts.

## ACKNOWLEDGEMENTS

This research was partially funded by FONDECYT ANID Grant No. 1231584. Part of this work was conducted while the author was a Visiting Professor under a Fulbright Chair at the Institute of Transportation Studies at the University of California, Davis. The author gratefully acknowledges valuable comments from David Bunch and Moshe Ben-Akiva, from two anonymous referees, as well as from participants at ICMC 2024 and IATBR 2024 on a preliminary version of this work. Any remaining errors are the author's responsibility.

## Declaration of generative AI and AI-assisted technologies in the manuscript preparation process

During the preparation of this work, the author used Claude.ai for English language editing/proofreading. The author reviewed and edited the output as needed and take full responsibility for the content of the published article.

## Appendix

**Proposition 2**: **Unbiasedness and Consistency of the Choice Frequency Estimator**

### 1. Setup and Definitions

We assume a Logit data-generating process (DGP) where an individual $n$ makes R+1 choices from a latent consideration set $C_n$.

The true choice probability for alternative $i$ in $C_n$ is constant across all choice instances $r = 1, \ldots, R+1$ (Homogeneity Assumption):

$$P_n\left(i \mid C_n\right)=\frac{e^{\mu V_{in}}}{\sum_{j\in C_n} e^{\mu V_{jn}}}$$

Let $y_{jn}^{(r)}$ be an indicator variable equal to 1 if alternative $j$ is chosen in instance $r$, and o otherwise.

The count $m_i$ represents the total number of times alternative $i$ is chosen across the $R$+1 instances:

$$m_i=\sum_{r=1}^{R+1} y_{in}^{(r)}$$

The estimator for the probability is defined as the empirical frequency:

$$\hat{P}_n\left(i \mid C_n\right)=\frac{m_i}{\sum_{j\in D_n} m_j}=\frac{m_i}{R+1}$$

### 2. Proof of Unbiasedness

Under the homogeneity assumption, each choice instance is a trial with probability $P_n\left(i \mid C_n\right)$. The expectation of the sum of indicators is the sum of their individual expectations:

$$E\left[m_i\right]=E\left[\sum_{r=1}^{R+1} y_{in}^{(r)}\right]=\sum_{r=1}^{R+1} E\left[y_{in}^{(r)}\right]=\left(R+1\right)\left(1P_n\left(i \mid C_n\right)+0\left(1-P_n\left(i \mid C_n\right)\right)\right)=\left(R+1\right)P_n\left(i \mid C_n\right)$$

Thus, the expectation of the estimator is

$$E\left[\hat{P}_n\left(i \mid C_n\right)\right]=\frac{E\left[m_i\right]}{R+1}=\frac{\left(R+1\right)P_n\left(i \mid C_n\right)}{R+1}=P_n\left(i \mid C_n\right)$$

**Conclusion:** The estimator is unbiased.

**3. Proof of Consistency (R →∞)**

To establish consistency, we examine the variance of the estimator. Since choices are assumed to be independent across instances (i.i.d. errors), the variance of the sum $m_i$ follows the properties of a binomial distribution (within the multinomial vector *M*:

$$Var\left[m_i\right] = Var\left[\sum_{r=1}^{R+1} y_{in}^{(r)}\right] = \sum_{r=1}^{R+1} E\left[y_{in}^{(r)}\right] = \left(R+1\right) P_n\left(i \mid C_n\right)\left(1 - P_n\left(i \mid C_n\right)\right)$$

The variance of the estimator $\hat{P}_n$ is then:

$$Var\left[\hat{P}_n\left(i \mid C_n\right)\right] = Var\left[\frac{m_i}{R+1}\right] = \frac{\left(R+1\right) P_n\left(i \mid C_n\right)\left(1 - P_n\left(i \mid C_n\right)\right)}{\left(R+1\right)^2} = \frac{P_n\left(i \mid C_n\right)\left(1 - P_n\left(i \mid C_n\right)\right)}{R+1}$$

As the number of historical instances **R** →∞, the variance collapses to zero:

$$\lim_{R \to \infty} Var\left[\hat{P}_n\left(i \mid C_n\right)\right] = 0 .$$

By Chebyshev's Inequality, because the estimator is unbiased and its variance vanishes as *R* increases, the estimator $\hat{P}_n\left(i \mid C_n\right)$ converges in probability to the true value $P_n\left(i \mid C_n\right)$.

**Conclusion:** The estimator is consistent as **R** →∞.

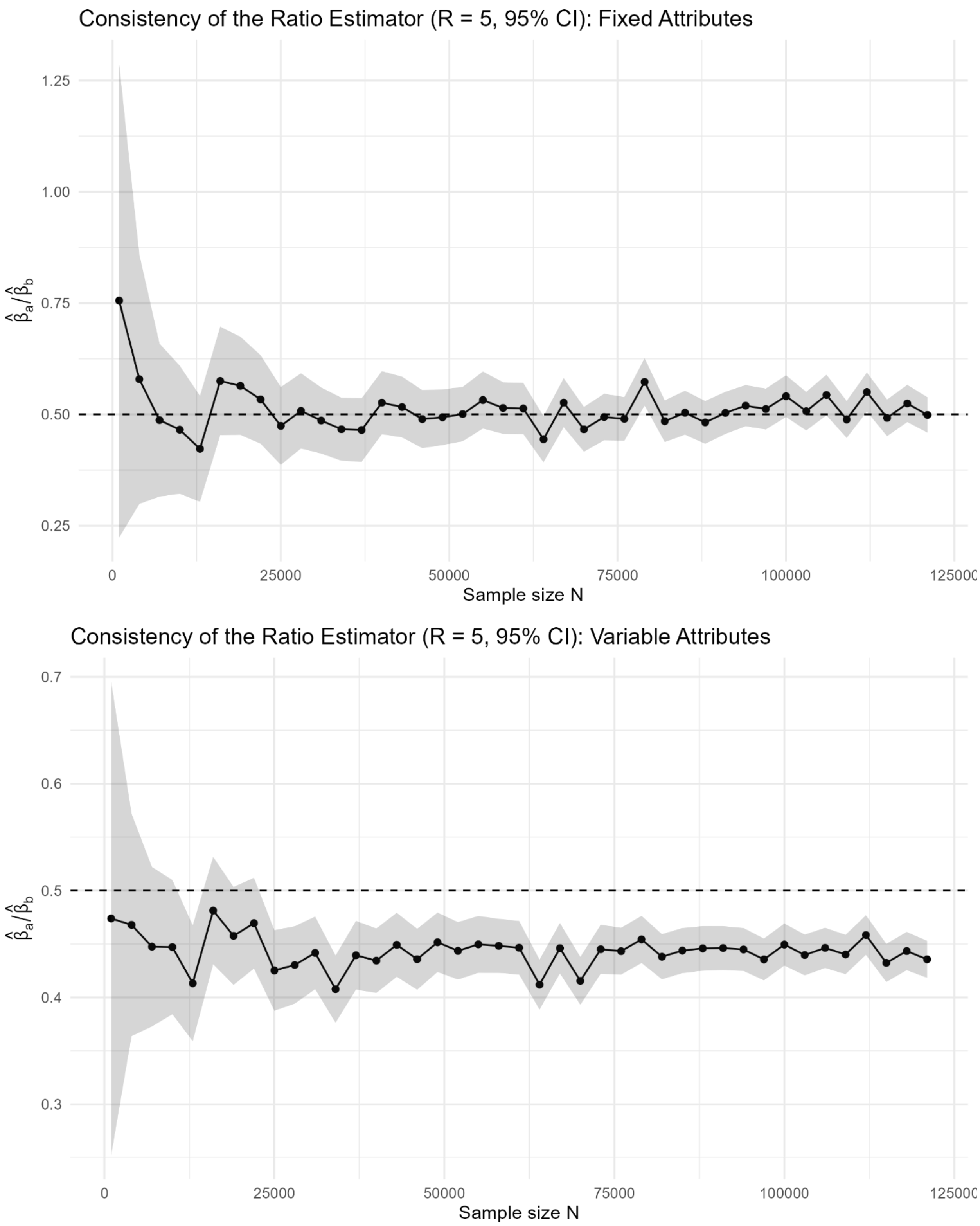

Figure 2. Analysis of Empirical Consistency For R=5

Table 3. Detailed Parameter Estimates For Different R. RMSE and Coverage

| | | Fixed | | | | Variable | | | |
|---|---|---|---|---|---|---|---|---|---|
| R | parameter | Mean_Est | Mean_SE | RMSE | Coverage | Mean_Est | Std_Error | RMSE | Coverage |
| 5 | $\hat{\beta}_a$ | 0.185 | 0.069 | 0.324 | 1% | 0.436 | 0.089 | 0.109 | 87% |
| 5 | $\hat{\beta}_b$ | 0.371 | 0.075 | 0.634 | 0% | 1.000 | 0.098 | 0.097 | 94% |
| 5 | $\hat{\beta}_p$ | -0.750 | 0.082 | 1.253 | 0% | -1.997 | 0.100 | 0.099 | 93% |
| 10 | $\hat{\beta}_a$ | 0.284 | 0.066 | 0.227 | 11% | 0.451 | 0.078 | 0.092 | 87% |
| 10 | $\hat{\beta}_b$ | 0.569 | 0.073 | 0.438 | 0% | 0.998 | 0.088 | 0.088 | 92% |
| 10 | $\hat{\beta}_p$ | -1.143 | 0.081 | 0.861 | 0% | -1.997 | 0.080 | 0.080 | 95% |
| 20 | $\hat{\beta}_a$ | 0.367 | 0.065 | 0.150 | 48% | 0.471 | 0.071 | 0.076 | 93% |
| 20 | $\hat{\beta}_b$ | 0.737 | 0.072 | 0.274 | 6% | 0.997 | 0.083 | 0.083 | 91% |
| 20 | $\hat{\beta}_p$ | -1.478 | 0.080 | 0.529 | 0% | -1.997 | 0.075 | 0.075 | 94% |
| 40 | $\hat{\beta}_a$ | 0.425 | 0.065 | 0.102 | 74% | 0.490 | 0.069 | 0.069 | 95% |
| 40 | $\hat{\beta}_b$ | 0.853 | 0.071 | 0.167 | 50% | 0.996 | 0.078 | 0.077 | 94% |
| 40 | $\hat{\beta}_p$ | -1.709 | 0.078 | 0.302 | 4% | -1.997 | 0.071 | 0.071 | 96% |